\documentclass{osa-article}

\journal{oe}

\usepackage{lipsum}
\usepackage{physics}

\DeclareUnicodeCharacter{0303}{\'{e}}
\DeclareUnicodeCharacter{0301}{\~{n}}
\usepackage{siunitx}
\newcommand{\dint}{D_\mathrm{int}}

\newcommand{\um}{\si{\um}~}

\begin{document}
	
\title{Frequency correlated photon generation at telecom band using silicon nitride ring cavities}
	
\author{Zhenghao Yin,\authormark{1,3} Kenta Sugiura,\authormark{1,3} Hideaki Takashima,\authormark{1} Ryo Okamoto,\authormark{1} Feng Qiu,\authormark{2} Shiyoshi Yokoyama,\authormark{2} and Shigeki Takeuchi\authormark{1,*}}
\address{\authormark{1}Department of Electronic Science and Engineering, Kyoto University, Kyotodaigakukatsura, Nishikyo-ku, Kyoto 615-8510, Japan\\	
\authormark{2}Institute for Materials Chemistry and Engineering, Kyushu University, 6-1 Kasuga-koen,
Kasuga-city, Fukuoka 816-8580, Japan \\
\authormark{3}These authors contributed equally to this work.
}
	\email{\authormark{*}takeuchi@kuee.kyoto-u.ac.jp} 
	
	
	\begin{abstract}
	
    Frequency entangled photon sources are in high demand in a variety of optical quantum technologies, including quantum key distribution, cluster state quantum computation and quantum metrology. In the recent decade, chip-scale entangled photon sources have been developed using silicon platforms, offering robustness, large scalability and CMOS technology compatibility.
    Here, we report the generation of frequency correlated photon pairs using a 150-GHz silicon nitride ring cavity. First, the device is characterized for studying the phase matching condition during spontaneous four-wave mixing. Next, we evaluate the joint spectrum intensity of the generated photons and confirm the photon pair generation in a total of 42 correlated frequency mode pairs, corresponding to a bandwidth of 51.25 nm. Finally, the experimental results are analyzed and the joint spectral intensity is quantified in terms of the phase matching condition.
    \end{abstract}
	
	\section{Introduction}
	
	The field of quantum information processing technology using photons has been growing in last two decades, including quantum cryptography \cite{Weedbrook2012,Scarani2009},
	quantum sensing \cite{Giovannetti2011, Degen2017}
	and linear optical quantum computation \cite{Kok2007, Ono2017, Okamoto2011}.
	For these applications, entangled photon sources \cite{Takeuchi2014} are useful for the generation of heralded single photons\cite{Kiyohara2020} and also as a resource for beating the quantum limit for phase measurement using an optical interferometer\cite{Nagata2007, Okamoto2008}.
	Especially, frequency entangled photons are attracting attention for realizing high-dimensional quantum entangled states\cite{Kues2017b, Reimer2019}. 
    They can also be used for quantum optical coherence tomography (QOCT) \cite{Abouraddy2002,Okano2015} and dense quantum key distribution \cite{Wengerowsky2018}. 

	The broad bandwidth of the frequency entangled photons is quite important for these applications. The depth resolution of QOCT image is given by the inverse of the bandwidth. Thus the broader the bandwidth of frequency entangled photons, the better the depth resolution of QOCT. Another example is the two photon absorption using entangled photons \cite{Dayan2004}. The larger bandwidth means the shorter correlation time of the bi-photons, resulting in the larger enhancement of TPA using frequency entangled photons. In addition, the number of frequency correlated modes can be increased using a broadband photon pair source for a given mode spacing. 

	Ultra-broad frequency entangled photons have been successfully generated using chirped quasi-phase matched devices\cite{Sensarn2010,Tanaka2012}. As an alternative approach, on-chip frequency-entangled photon sources realized by CMOS compatible platforms are attractive in terms of system scalability and initialization. Frequency entangled photon generation using silicon waveguides has been reported. However, for two-photon absorption in silicon, the flux of generated photon pairs is limited. To overcome this limitation, on-chip ring resonators using high-index contrast doped glass (HICDG) \cite{Razzari2010,Reimer2014a,Ambrosetti2016,Kues2017b,Roztocki2017,Sugiura2019,Sugiura2020} and silicon nitride (SiN) \cite{Moss2013,Ramelow2015a} have been studied. In Ref. \cite{Ambrosetti2016}, the bandwidth of 140 nm has been observed in the single-photon spectrum and the frequency correlation has been measured for the bandwidth of 10 nm. However, the confirmed bandwidth of the frequency-correlated photon pair generation has, to date, been limited to 16 nm using a SiN ring device\cite{Ramelow2015a} and to 23.6 nm using a HICDG ring device\cite{Sugiura2020}, which is the largest bandwidth to the best of our knowledge.

    In this paper, we report the broadband generation of photon pairs using a SiN ring resonator. By adopting a device structure in which the material dispersion is well compensated by the structural dispersion, we obtain very small dispersion for a broad wavelength range. From the obtained joint spectrum using a frequency-resolved coincidence measurement using superconducting nanowire single photon detectors, we confirm photon pair generation correlated in frequency over a 51-nm range, which is more than two times broader than the previous record of the confirmed bandwidth of the frequency correlated photon pairs\cite{Sugiura2020}. Furthermore, the observed joint-spectral intensities are well reproduced by the theoretical calculation.
    
    In the following, we describe the structure of the device in section 2. In section 3, a characterization of the device using a transmission spectrum is provided. In sections 4 and 5, the experimental setup and the results of broad-band frequency-correlated photon pair generation are explained and discussed.
    
    \section{On-chip silicon nitride ring resonator}
	In a silicon nitride ring cavity pumped with a continuous wave laser,
	signal and idler photon pairs arise simultaneously due to the perturbation of Kerr nonlinearity.
	To increase the frequency correlation range of generated photon pairs, the phase matching condition is satisfied in a broadband form, in both the frequency domain and momentum domain \cite{Chen2011}. 
	This is a challenge for device fabrication since the cavity dimension affects the frequency dispersion significantly.
	
\begin{figure}[t]
		\centering
		\includegraphics[width=3in]{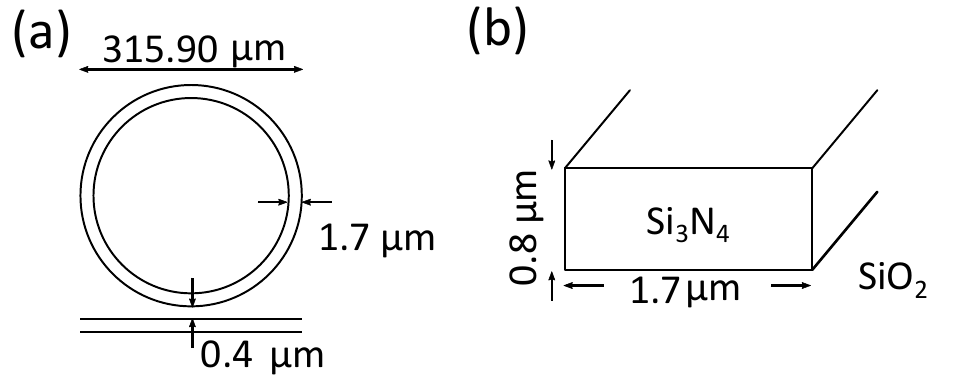}
		\caption{(a) Top view of the silicon nitride ring resonator used in this study. The diameter corresponds to a near-150-GHz mode spacing. (b) 1.7 \um $\times$ 0.8 \um cross section of ring cavity, enabling anomalous dispersion behavior.}
		\label{fig:device}
	\end{figure}

	Over the last decade, the fabrication process based on silicon nitride material with a Kerr frequency comb and optical soliton generation has improved, especially for high quality factor ring cavities\cite{Moss2013, Cheng2017b}.
	In our study, we customized silicon nitride ring cavities via the photonic damascene process\cite{Pfeiffer2017, Pfeiffer2018a} to obtain a low anomalous group velocity dispersion.
	To cope with both small dispersion and the high $Q$ factor, we made ring resonators with 5 different ring sizes with 4 different gap distances between the ring and the bath line, and selected the ring diameter of 314.90 \um and the gap distance of 0.4 \um.
	The device is illustrated in \hyperref[fig:device]{Fig.~1}. The ring diameter 
	corresponds to the 150-GHz free spectra range (FSR), which corresponds to the spacing of the neighboring frequency-correlated photon pair mode.
	The widths of the bus waveguide and ring waveguide are both 1.7 \um. 
	The thickness of the silicon nitride layer is 800 nm, and the layer is surrounded with silicon dioxide as a low-index cladding layer.
	
	\section{Device characterization}
	
	\subsection{Device transmission}
		
	\begin{figure}[!t]		\centering
		\includegraphics[width=4.3in]{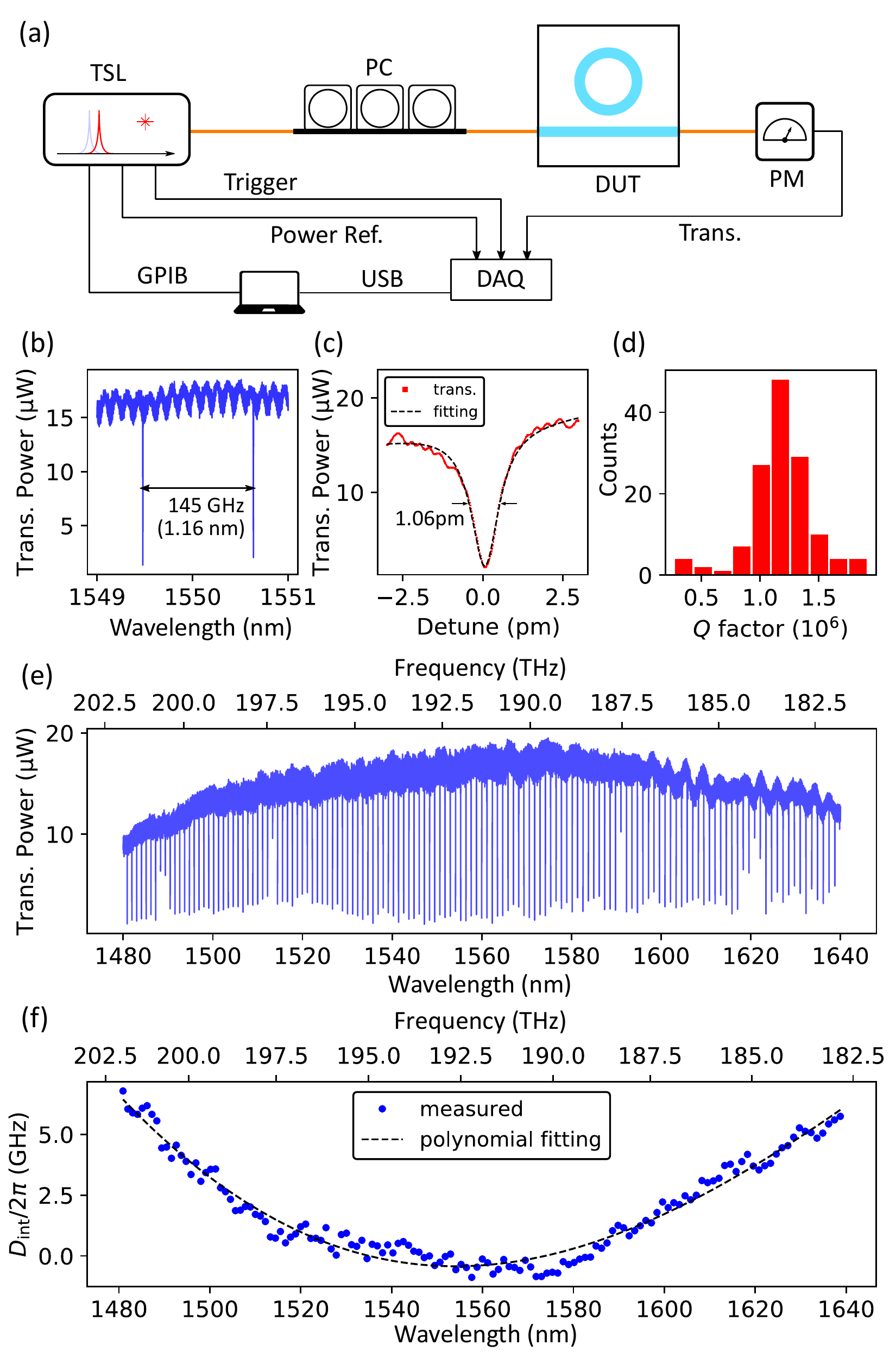}
		\caption{Silicon nitride cavity for generating frequency-correlated photon pairs. (a) Schematic diagram of device transmission measurement. TSL: tunable semiconductor laser, PC: polarization controller, PM: power meter, DAQ: data acquisition device. The orange lines are single mode single mode optical fibers and the black lines are electric lines. (b) Free spectral range of device near 1550 nm. (c) Resonance dip at 1550.64 nm, indicating a \textit{Q}-factor of \num{1.46d6}. (d) \textit{Q}-factors counts, up to \num{d6} order for almost all the resonances. (e) Device transmission from 1480 nm to 1640 nm. (f) Integrated dispersion extracted from the device transmission and calculated based on the resonant wavelengths.}
		\label{fig:trans}
	\end{figure}
	To study the device performance and phase matching condition, we first measure the device transmission using the setup shown in \hyperref[fig:trans]{Fig.~2(a)}. 
	A tunable semiconductor laser (TSL, santec TSL-710) is used to sweep the wavelength from 1480 nm to 1640 nm.
	After aligning the input mode to launch only the fundamental TE mode using a fiber polarization controller (PC), the device is coupled with two lensed fibers, whose typical facet-to-facet loss is around $-6$ dB in our experiments and placed over a thermoelectric cooler (TEC) to improve the thermal stability. 
	The transmitted power from the device is then collected by a power meter (PM, Newport, 2936-C) and finally acquired by a data acquisition module as well as the laser sweeping trigger signal and power reference. 
	
	The transmission of this device near 1550 nm is shown in \hyperref[fig:device]{Fig.~2(e)}. 
	A magnified view of the device transmission near 1550 nm is plotted in \hyperref[fig:device]{Fig.~2(b)}. The observed FSR is around 145 GHz near 1550 nm.
    The transmission spectrum of a single resonant dip is shown in \hyperref[fig:device]{Fig.~2(c)}. The dashed line is a Lorentzian fitting of the resonant dip, indicating a 1.06 pm width and a \textit{Q}-factor of \num{1.46d6}.
	The \textit{Q}-factors of all the resonant dips within the wavelength range between 1480 nm and 1640 nm are counted and presented in \hyperref[fig:device]{Fig.~2(d)}.
	The maximum quality factor is $1.71 \times 10^6$ and the mean quality factor is $1.07\times 10^6$.

	\subsection{Dispersion analysis}
	
	Next, we analyzed the device dispersion of the ring resonator.
	Following the integrated dispersion approach\cite{Brasch2014a}, we evaluated the second-order mode dispersion of the silicon nitride ring cavity.
	
	After the cavity resonant frequencies $ \omega_{\mu} $ are extracted from the device transmission, a Taylor series can be used to expand the resonant frequencies with respect to the relative mode index $\mu$, where $\mu=0$ corresponds to the pump mode during the SFWM.
	\begin{align}
		\dint(\mu) 
		&= \omega_{\mu} - (\omega_0 + \mu D_1)  \nonumber \\
		&= \frac{D_2}{2}\mu^2 + \frac{D_3}{6}\mu^3 + \cdots 
	\end{align}
	where $\dint$ is the integrated dispersion and $D_1$ is the FSR of the angular frequency.
	As shown in \hyperref[fig:trans]{Fig.~2(f)}, a polynomial fitting of the integrated dispersion of resonant frequency gives $D_2=0.71$ MHz and $D_3=6.37$ kHz,
	showing that the frequency mismatch between tens of modes is less than the width of the cavity resonant mode (132 MHz).

	\section{Broadband photon pair generation}
		
	\begin{figure}[!b]		\centering\includegraphics[width=5.5in]{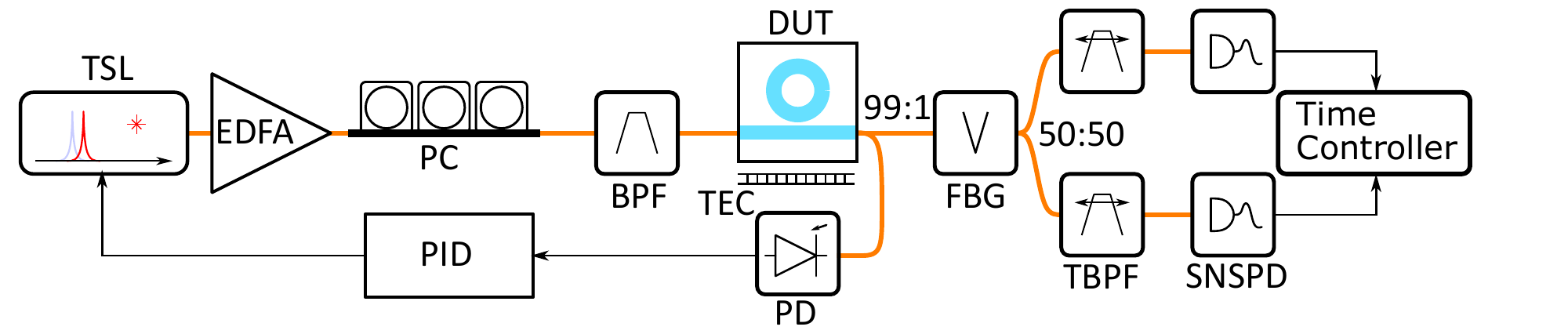}
		\label{fig:setup}
		\caption{Schematic diagram of mode-resolvable photon pair generation. TSL: tunable semiconductor laser, EDFA: erbium doped fiber amplifier, PC: polarization controller, BPF: bandpass filter, PD: photodetector, FBG: fiber Bragg grating filter, TBPF: tunable bandpass filter, SNSPD: superconducting nanowire single photon detector. The orange lines are optical fibers and the black lines are electric lines.}
	\end{figure}
	Finally, mode-resolvable photon pair generation is demonstrated using the setup shown in \hyperref[fig:setup]{Fig.~3}.  
	To increase the pump power in the ring cavity, an erbium doped fiber amplifier (EDFA, Alnair Labs, HPA-200C) is employed as well as the TSL.
	The input polarization is aligned using a polarization controller. 
	A bandpass filter is used to reject the sideband noise of the EDFA.
	The pump wavelength $\lambda_\mathrm{p}$ is set to 1550.63 nm to match the cavity resonant mode. The power measured before the device coupling is 24.5 mW.
	Considering the thermal instability during the pump wavelength location, a simple feedback loop is added in the scheme.
	At the device output port, a 99:1 beam splitter is connected for fine control of the pump wavelength to maintain resonance alignment.
	1\% of the device output is detected with a photodiode to monitor the ratio of the transmitted power to the input power, which is around $-10$ dB during our experiments. 
	The other 99\% of the output power is filtered to reject the pump light with a fiber Bragg grating (FBG, OE Land) and then split into signal and idler channels using another 50:50 beam splitter. 
	Tunable bandpass filters (TBPFs, WL Photonics) pass the needed signal and idler modes, with a band window of 0.12 nm, much narrower than the device mode spacing.
	Both channels are filtered to identify the signal or idler modes, and are finally detected with superconducting nanowire single photon detectors (SNSPDs, SCONTEL). 
	Finally, a time-to-digital converter (ID Quantique, id900) records all the signal and idler counting events to allow evaluation of the photon pair generation rate for specific mode pairs.
	In order to confirm the frequency correlation for large bandwidth, we realized a stable measurement system where the wavelength of the narrow-band CW pump laser was actively controlled by proportional-integral-differential (PID) controller with monitoring the output pump laser power from the device.
	
	To evaluate the photon pair generation rate, the losses of both the signal and idler channels are measured.
	The transmission of the lensed-fiber coupling is $-3.0$ dB per facet. The measured total transmission of FBG, TBPF and beamsplitter is about $-7.0$ dB.
	In the SNSPD, the detection efficiencies are 50\%. 
	Hence, the total detection efficiencies are $-13.0$ dB for both signal and idler photons in our experiments.

	\section{Results and discussion}

\begin{figure}[!b]
			\centering\includegraphics[width=4.2in]{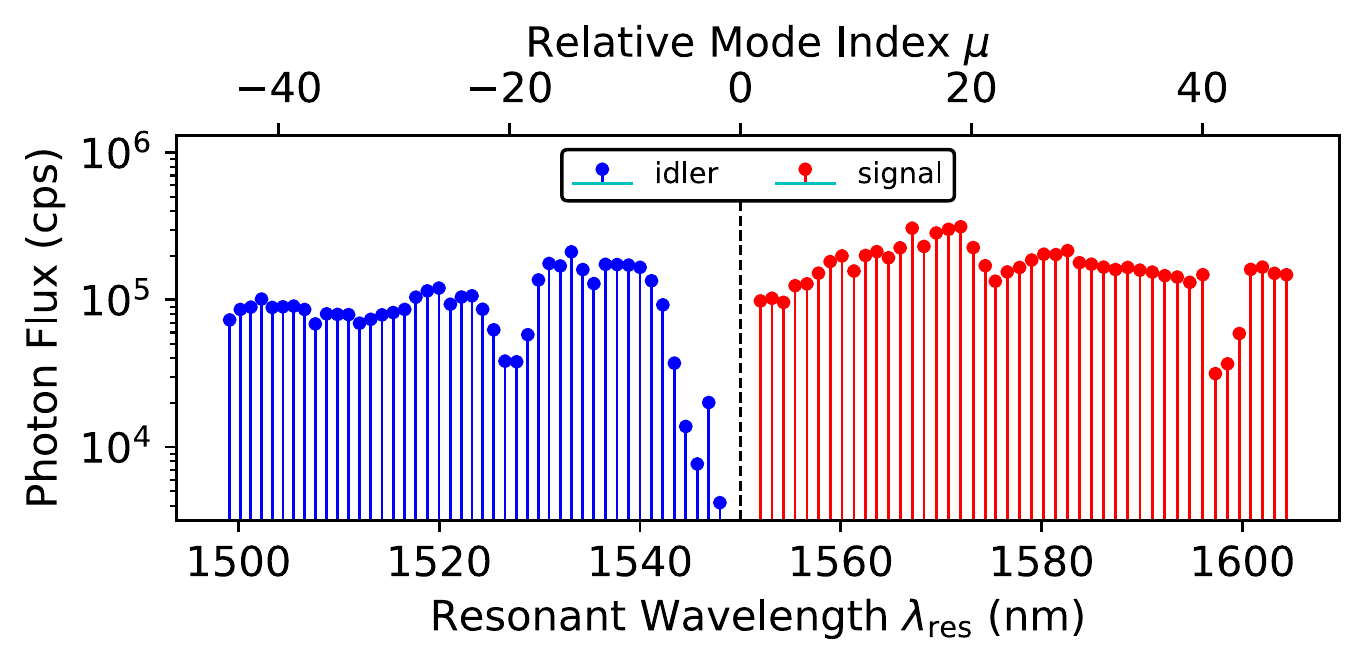}
			\caption{Photon flux of all the resonant modes filtered into signal and idler channels with an accumulation time of 10 s. Several modes around 1550 nm are affected by the FBG used to remove the pump light in our setup.}
			\label{fig:flux}
		\end{figure}

	\hyperref[fig:flux]{Fig.~4} shows the photon flux at cavity resonant modes, with red and blue corresponding to the signal and idler modes respectively.
	Here, mode 0 is set as the pump frequency. The photon flux of the $-8$th to $-1$st modes is significantly affected by the FBG used to stop the pump light. 
	The mean photon flux for all the modes is up to \num{1.7d5} cps. 
		\begin{figure}[t]		\centering\includegraphics[width=4.2in]{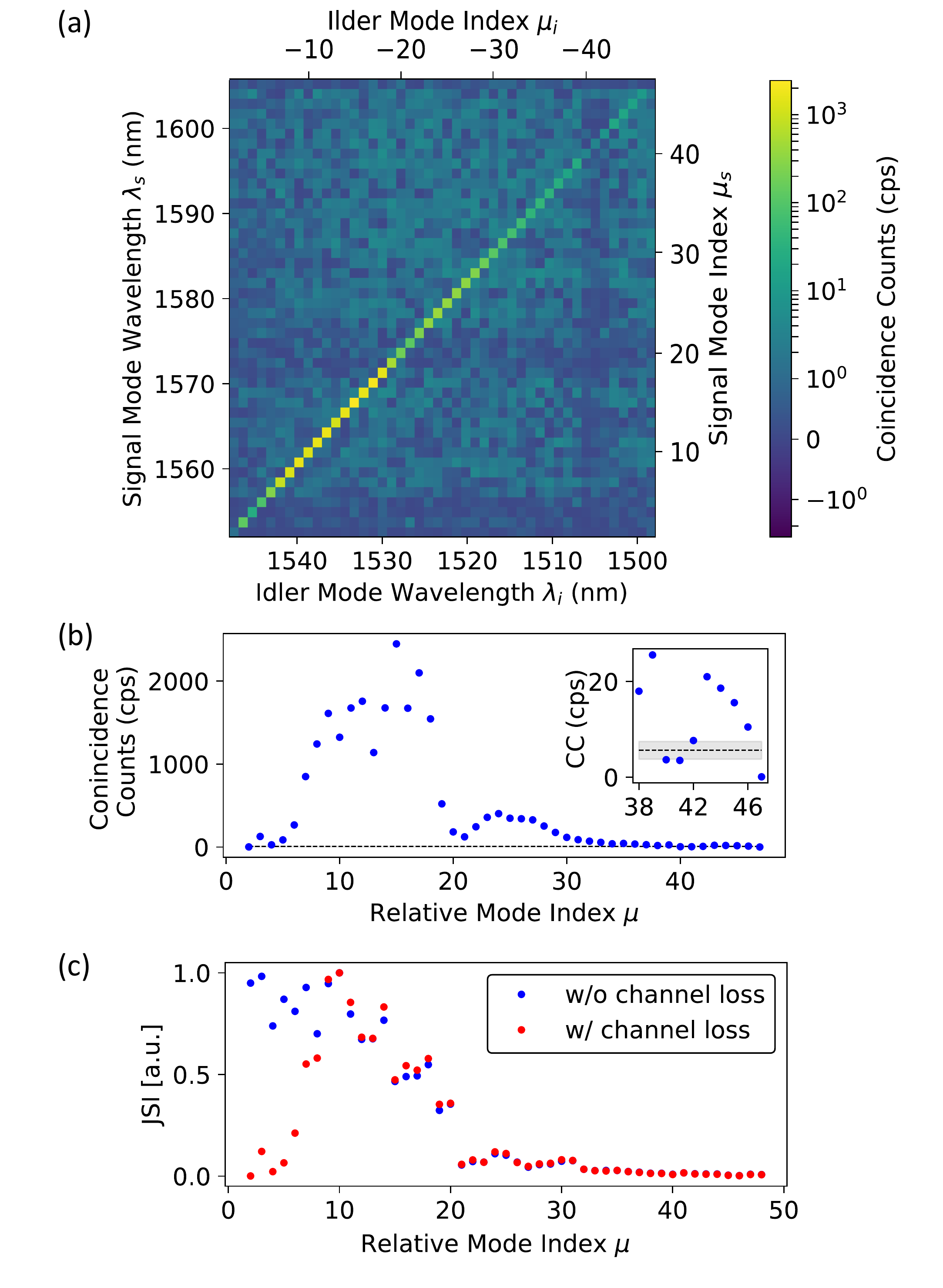}
		\caption{Result of mode resolvable photon pair generation. \textbf{a} Measured joint spectral intensity, including $46\times46$ mode pairs. The diagonal terms are replotted in \textbf{b} and are compared with a numerical analysis result in \textbf{c}. The insert in \textbf{b} shows the coincidence counts at the 40th and 41st mode pairs, which are lower than the maximal counting of off-diagonal terms.}
		\label{fig:result}
	\end{figure}
	
	Next, the joint spectral intensity (JSI) of generated photon pairs is evaluated by coincidence counting (CC) with a time window $t_c$ of 1 ns and an accumulation time of 10 s for each mode-by-mode photon count.
	As shown in \hyperref[fig:result]{Fig.~5(a)}, for the ranges from mode 1499.09 nm ($\mu=-47$) to 1548.32 nm ($\mu=-2$) in the idler band, and from 1552.95 nm ($\mu=2$) to 1605.79 nm ($\mu=47$) in the signal band, the JSI map covers all the $46\times46=2,116$ mode pairs.
	To reduce the effect of accidental counting events during the measurement, the accidental coincidence count (ACC), defined as $N_\mathrm{s} N_\mathrm{i} t_c $, is subtracted from the result.
	The mean ACC for all the modes is  10.5 cps, and the coincidence to accidental ratios are of diagonal terms from 2th mode to 45th mode are larger than 1. 
	Along the diagonal direction, the maximal CC is \num{2.45d3} cps and decreases significantly	as the relative mode index increases. This can be explained in terms of the dispersion of the device.
    As illustrated in \hyperref[fig:result]{Fig.~5(b)}, the highest value of the off-diagonal terms, i.e. non-phase matched modes, is 5.61$\pm$1.85 cps. 
	This is higher than the counts for the 2nd, 40th, 41st and 47th mode pairs.
	Therefore, we conclude that 37 continuous mode pairs and a total of 42 mode pairs are generated in our experiment, which corresponds to a bandwidth of 51.25 nm in the signal band. Note that the full bandwidth of the photon-pairs from the $-46$th mode to 46th mode is 105.25 nm.

	To clarify how the dispersion dominates the broadband SFWM process, we calculate the JSI of photons in the nonlinear cavity. Based on the resonant mode lineshape function and phase mismatch, the probability of two-photon state generation can be written as \cite{Sugiura2020}
	\begin{align}\label{eq:jsi}
	C_\mathrm{JSI}(\mu)  
	&= \eta_{\mu} \eta_{-\mu}  \int_{-\infty}^{\infty}  \dd \omega_\mathrm{p}  g(\omega_\mathrm{p}) 
	\int_{-\infty}^{\infty}  \dd \Omega  
	A^2_{\mu}(\omega_\mathrm{p} + \Omega) A^2_{-\mu}(\omega_\mathrm{p} - \Omega) 
	\mathrm{sinc}^2  (\Delta\phi)
	\end{align}
	where $ \eta_{\mu, -\mu} $ is a constant of the cavity mode, including the detection channel loss coefficient; $ g(\omega_\mathrm{p}) $ is the pump light lineshape function; 
	$ \Omega $ is the frequency deviation from pump wavelength; 
	$ A_{\mu}(\omega) $ is the cavity transmission given by a Lorentzian function:
	\begin{equation}\label{eq:loren}
	    A_{\mu}(\omega) = \frac{\sqrt{\gamma_{\mu}}}{\gamma_{\mu}/2 - {\rm i}(\omega-\omega_{\mu})}
	\end{equation} 
	where $\gamma_{\mu}$ is the full width at half maximum (FWHM);  and  
	\begin{align}\label{eq:dphi}
	  \Delta\phi 
	  &= (\omega_\mathrm{s} + \omega_\mathrm{i} - 2 \omega_\mathrm{p})\tau + \Gamma P_\mathrm{p}   \nonumber \\
	  &= \dint(\mu)\tau +\dint(-\mu)\tau+\Gamma P_\mathrm{p}    
	\end{align}
	is the mismatch due to the frequency mismatch and power-dependent self-phase modulation during four wave mixing phase, where $\dint(\mu)$ is the mode integrated dispersion,
	$ \tau $ is the cavity round-trip time,
	$\Gamma$ is the Kerr nonlinear coefficient and $ P_\mathrm{p} $ is the intracavity power of the pump light.
	
	Thus, the diagonal terms in the JSI map can be evaluated numerically by substituting the retrieved cavity resonant frequencies and linewidths into \hyperref[eq:jsi]{Eq. 1}.
    The result is presented in \hyperref[fig:result]{Fig.~5(c)} and compared with the experimental CC result in \hyperref[fig:result]{Fig.~5(b)}. 
    It can be seen that the numerical analysis agrees well with the experimental result of CC and thus validates the phase matching theory of SFWM.
    As shown in Fig. 5 (b) and 5 (c), the CCs around the 20th mode are significantly low. 
    This can be explained by the term $\int_{-\infty}^{\infty}  \dd \Omega  
	A^2_{\mu}(\omega_\mathrm{p} + \Omega) A^2_{-\mu}(\omega_\mathrm{p} - \Omega) $ in \hyperref[eq:jsi]{Eq. (2)}, which relates the resonant frequencies and energy conservation of SFWM.  The frequency mismatch between pump photons ($2\omega_{\rm p}$) and signal photon and idler photons ($\omega_\mu + \omega_{-\mu}$) increases around the 20th mode.

	\section{Conclusion}
	In conclusion, we realized broadband frequency-correlated photon generation using a silicon nitride ring cavity. From the experimentally obtained JSI, we confirmed that the generated photon pairs are correlated in 37 continuous frequency mode pairs and a total of 42 mode pairs, which corresponds to a bandwidth of 51.25 nm in the signal band. Note that the full bandwidth of the photon-pairs from the $-46$th mode to 46th mode is 105.25 nm. The cavity is characterized for studying the phase matching condition. We then demonstrated broadband photon pair generation using 24.5 mW pump power.
	With these obtained data, we have successfully reconstructed the intensities of the diagonal part of JSI using the estimated dispersion without any fitting parameters.
	We anticipate that the on-chip scale nonlinear ring cavity will boost the development of scalable frequency-entangled photon sources and enable future frequency-encoded quantum technology.
	
	\section*{Funding}
	JST-CREST (JPMJCR1674); MEXT Q-LEAP (JPMXS0118067634); JSPS-KAKENHI (26220712); Grant-in-Aid for JSPS Research Fellow (19J20968); MEXT WISE Program; Research Program for Next Generation Young Scientists of ``Five-star Alliance" in ``NJRC mater. \&  Dev.''.
	
	\section*{Acknowledgments}
	We wish to acknowledge the helpful comments provided by Bo Cao, Takayuki Kiyohara and Xiaoyang Cheng.
	
	\section*{Disclosures}
	The authors declare no conflicts of interest.
	

\end{document}